\documentstyle[prd,tighten,aps]{revtex}
\begin{document}
\tightenlines
\title{Natural units, numbers and numerical clusters}
\author{F. Pisano and N. O. Reis}
\address{Department of Physics, Federal University of Paran\'a, Curitiba, 
PR 81531-990, Brazil}
\maketitle
\bigskip
\begin{abstract}
Defined by Lord Kelvin as the science of measurement it is described a   
fundamental fact of physics. The so called `natural' units represent the  
unique system of units conveniently used in the realm of High Energy
Physics.
The system of natural units is defined by the consistent providing of
the adimensional unit value to the velocity $c$ of the eletromagnetic
radiation in the vacuum, related to the Maxwell classical electromagnetic 
theory and to the special relativity theory, to the reduced Planck
constant $\hbar\equiv\frac{h}{2\pi}$ having the dimension of action,
which is the fundamental constant of the quantum
mechanics as well as to the Boltzmann constant $k_{\rm B}$ which has the
dimension  of heat capacity, being the fundamental constant associated to
the statistical mechanics, $c=\hbar = k_{\rm B} = 1$. Such prescription is
mandatory as we investigate the ultimate constituents of matter and their
dynamics, since the atoms, leptons and quarks up to the grand unification,
strings, superstrings, and the supergravity theories, as well as the total
unification of the interactions in the M-theory. The one-dimensional
character
of this phenomenology is totally featured by the use of the natural units,
reflecting the fact that in order to be possible the investigation of the 
fundamental constituents of matter it is being necessary the resolution of
lengths  of the order of $10^{-16}\,\,{\rm cm}$ or to reach energies of
the order of at least 1 TeV. The application of this system of natural
units results
exceptionally adequate in the description of the Universe since the Planck
scale, up to the Hubble large scale
expanding Universe, resulting in the curious formation of the regular
numerical clusters.
\end{abstract}
06.20.Fn: Units and standards \\
06.30.-k: Measurements common to several branches of physics and astronomy
\twocolumn[\hsize\textwidth\columnwidth\hsize\csname@twocolumnfalse\endcsname
\vskip2pc]
\section{Introduction}
\par
The consolidation of specific systems of units of measurement was determined 
by factors related to the environment, as well as by historical and 
physiological factors. Such fact is possible to be verifyed for instance 
through the case of the units inch, yard, metrical foot, palm and so on. 
Considering such natural 
arbitrariety in the definition of the patterns of measurement it was always 
necessary to try to establish systems of simple and convenient units such as, 
for instance, the case of the m.k.s. system, the base of the 
international system of units, or even the case of the c.g.s. system.
Naturally, the choice of such standards is arbitrary. Nevertheless, the m.k.s. 
or c.g.s. systems has been becoming commonly used in the realm of physics 
and engineering. In the High Energy Physics the system of natural units 
is the exceptionally consistent system 
not requiring the definitions of the usual patterns of measurement, 
generically used, being possible the definition of new fundamental
patterns of measurement. The strongly singular but universal 
postulate used in this case consists in the fact that the unique things 
that can retain their identities 
in any place of the Universe are the elementary 
particles, leptons and quarks, as well as their interactions described by the 
corresponding dynamics given by the general relativity~\cite{TB} 
and by the standard model~\cite{WSG} 
of the nongravitational interactions, which consist in 
the weak and strong nuclear interactions of short range, $10^{-16}$ cm for 
the weak interaction and $10^{-13}$ cm for the strong 
interaction, and the long range electromagnetic interaction. In such physics 
where we are searching the fundamental laws of the Universe as a whole 
and the bits of matter, measurement is essential again, but the quantum and 
relativistic regimes imply, aside from any multicultural 
diversity~\cite{Romer} the simplest system of units, the `natural' units. 
\section{The Physical Constants and The Natural Units}
\par
As the description of the dynamics of the fudamental interactions was being
elaborated it became a consolidated fact that the universal physical 
constants usually play the role of constants of proportionality between 
magnitudes in the equations of physics. 
The velocity of the electromagnetic radiation in the vacuum,
\begin{equation}
c = 299\,792\,458\,\,{\rm m}\,{\rm s}^{-1}
\label{uno}
\end{equation}
which is an exact value since the meter turn to be defined as the space 
covered by the light in the vacuum during the time of 
1/299 792 458 of 1 second, the reduced Planck constant~\cite{PDG2000}
\begin{eqnarray}
\hbar \equiv \frac{h}{2\pi} & = & 
\frac{6.626\,068\,72}{2\pi}\times 10^{-34}\,\,
{\rm J}\,\,{\rm s} \nonumber \\
& = & 1.054\,571\,596\times 10^{-34}\,\,{\rm J}\,\,{\rm s}
\label{due}
\end{eqnarray}
and the Boltzmann constant,
\begin{eqnarray}
k_{\rm B} & = & 1.380\,650\,3\times 10^{-23}\,\,{\rm J}\,\,{\rm K}^{-1} 
\nonumber \\
& = & 8.617\,342\times 10^{-5}\,\,{\rm eV}\,\,{\rm K}^{-1},
\label{tree}
\end{eqnarray}
define the system of natural units when are satisfied the dimensional 
relations
\begin{equation}
[c] = [\hbar] = [k_{\rm B}] = 1
\label{quatr}
\end{equation} 
as well as, numerically, also\footnote{Notice the coincidence with the 
dimensional value $c = 1\,\,{\rm ly}\times{\rm yr}^{-1}$.}
\begin{equation}
c = \hbar = k_{\rm B} = 1.
\label{cinque}
\end{equation}
These three universal physical constants are associated with the 
electromagnetic Maxwell theory, as well as with the special relativity, $c$, 
with the quantum 
mechanics, $\hbar$, and with the statistical mechanics, $k_{\rm B}$, 
already appearing in the Boltzmann equation of calorimetry
\begin{equation}
\Delta Q = \Delta Q(\Theta) = k_{\rm B}\,\Delta\Theta
\label{sei}
\end{equation}
which establishes the relation between the perceptible heat quantity 
$\Delta Q$ and the finite variation $\Delta\Theta$ of the absolute 
temperature $\Theta$. 
The Newton-Cavendish constant of universal gravitation, by its turn,
\begin{equation}
G = 6.672\,599\times 10^{-11}\,\,{\rm kg}^{-1}\,\,{\rm m}^3\,\,{\rm s}^{-2}
\label{sette}
\end{equation}
is the constant appearing both in the Newtonian gravitation and in the 
general relativity. 
\par
Regarding the c.g.s. or m.k.s. systems the fundamental dimensions 
are length $[L]$, mass $[M]$, and time $[T]$ but taking the mass dimension 
$[M]$, action $[S]$, and velocity $[V]$ as fundamental dimensions, 
length and time dimensions are 
\begin{mathletters}
\begin{eqnarray}
[L] & = & \frac{[S]}{[M][V]} \\ 
\label{exuno} 
[T] & = & \frac{[S]}{[M][V]^2}
\label{exdue}
\end{eqnarray}
\end{mathletters}
and being $p$, $q$, $r$ real numbers the general dimensional relation 
\begin{equation}
[M]^p\,[L]^q\,[T]^r = [M]^{p-q-r}\,[S]^{q+r}\,[V]^{-q-2r}
\label{ottto}
\end{equation}
in the c.g.s. or m.k.s. units has the natural unit mass dimension $[M]^n$ 
with $n=p-q-r$. For action $p=1$, $q=2$, $r=-1$, and for velocity we must have 
$p=0$, $q=1$, $r=-1$ with $n=0$ for both of such magnitudes. The mass or 
energy, length, and time have the $p=1$, $q=r=0$, and 
$q=1$, $p=r=0$, and also the $r=1$, $p=q=0$ attributions with the 
respective mass 
natural unit dimensions $n=+1,-1,-1$. 
For any physical magnitude $A$ with the c.g.s. or m.k.s. units
\begin{equation}
[A] = [M]^p \, [L]^q \, [T]^r
\label{ottone}
\end{equation}
and if we choose as fundamental unit the mass $[M]$ the natural units (NU) 
and the International System of units, SI, from the French terminology 
{\it Syst\`eme International d'Unit\'ees}, of the $A$ quantity are related 
according to
\begin{equation}
[M]^\alpha\equiv [A]_{\rm NU} = [A]_{\rm SI}\, [\hbar ]^x\, [c]^y  
\label{ottwo}
\end{equation}
where $[\hbar]=[M]\,[L]^2\, [T]^{-1}$ and 
$[c]=[M]^0\, [L]\, [T]^{-1}$. Taking now $[A]$ writen 
as in Eq.~(\ref{ottone}) with the $[\hbar]$ and $[c]$ dimensional relations 
the Eq.~(\ref{ottwo}) becomes
\begin{equation}
[M]^\alpha = [M]^{p+x}\,[L]^{q+2x+y}\,[T]^{r-x-y}.
\label{ottunaltr}
\end{equation}
The values of $\alpha$, $x$, and $y$ are obtained by realizing that two 
quantities are equal just when the potencies of the three fundamental 
units are the same. Therefore, 
\begin{eqnarray}
[M]^\alpha & = & [M]^{p+x}, 
\nonumber \\
\left[L\right]^0 & = & [L]^{q+2x+y},\\
\label{upisz}
[T]^0 & = & [T]^{r-x-y}, 
\nonumber
\end{eqnarray}
and to know $\alpha$, $x$ and $y$ in terms of the $p$, $q$, and $r$ 
dimensional exponents it is just necessary to solve the system of algebraic 
equations,
\begin{eqnarray}
p + x & = & \alpha,
\nonumber \\
q + 2x + y & = & 0,
\label{unlprt}
\\
r - x - y & = & 0,
\nonumber
\end{eqnarray}
resulting
\begin{eqnarray}
\alpha & = & p - q - r,
\nonumber \\
x & = & -q-r, 
\label{unripet}
\\
y & = & q + 2r,
\nonumber
\end{eqnarray}
and we can write the dimension of the quantity $A$ in a mass natural unit as
\begin{mathletters}
\begin{equation}
[A]_{\rm NU} = [M]^{p-q-r}
\label{newuno}
\end{equation}
with the conversion factor
\begin{equation}
[A]_{\rm NU} = [A]_{\rm SI}\,(\hbar^{-q-r} c^{q+2r})
\label{nevdue}
\end{equation}
\end{mathletters}
given in terms of the $\hbar$ and $c$ universal constants.\\
The same general procedure gives the following results, 
\begin{mathletters}
\begin{equation}
[A]_{\rm NU} = [L]^{-p+q+r},
\label{newtre}
\end{equation}
\begin{equation}
[A]_{\rm NU} = [A]_{\rm SI}\,(\hbar^{-p} c^{p+ r})
\label{nevquattr}
\end{equation}
\end{mathletters}
when the fundamental unit is a length, and 
\begin{mathletters}
\begin{equation}
[A]_{\rm NU} = [T]^{-p+q+r},
\label{newcinq}
\end{equation}
\begin{equation}
[A]_{\rm NU} = [A]_{\rm SI}\,(\hbar^{-p}\,c^{2p - q})
\label{newssei}
\end{equation}
\end{mathletters}
for the time as natural unit, which comply the linear algebra of 
fundamental units~\cite{Maksymowicz}. 
The conversion equations allows us to convert any physical quantity to NU, 
or back to the SI. The conversion is realized by setting explicitly 
the $[A]_{\rm SI}$ as every SI unit which it contains. 
The dimensional pattern of some fundamental physical magnitudes in natural 
units is
\begin{equation}
[E] = [M] = [\Theta] = [L]^{-1} = [T]^{-1}
\label{nup}
\end{equation}
which can be verified in the following sequence, \\
(i) $[E] = [M]:$ \\
    it follows from the mass-energy relation $E = M c^2$; \\
(ii) $[E] = [\Theta]:$ \\
     take the energy-temperature Boltzmann equation, 
     $E(\Theta)=Q(\Theta)=k_{\rm B}\Theta$ where $\Theta$ is the absolute 
     temperature; \\
(iii) $[E] = [L]^{-1}:$ \\
      from the energy-momentum relativistic equation for a massless particle, 
      $E = p c$, where $p = |{\bf p}|$ is the scalar momentum, and the de 
      Broglie relation $\lambda = h/p$, associating the wave character 
      ($\lambda$) of the matter ($p$), it follows that $\lambda = h c/E$ or 
      $E = h c/\lambda$; \\
(iv) $[E] = [T]^{-1}:$ \\
     being $[c] = [L][T]^{-1} = 1$ in natural units, then from 
     $[E] = [L]^{-1}$ we see that $[E] = [T]^{-1}$.
\par
All the diversity of the world\cite{Weisskopf} is found at different sizes, 
energy, temperature or any other magnitude. We find the nucleons and atoms 
from the Fermi length, $1\,\,{\rm fermi}\equiv 10^{-15}\,\,{\rm m}$, to the 
Angstr\"om scale, $1\,\,{\rm\AA}\equiv 10^{-10}\,\,{\rm m}$, virus between 
$(20\,\,{\rm and}\,\,300)\times 10^{-9}\,\,{\rm m}$, bacteria and cellule at 
$1\,\,\mu {\rm m}$. Atoms form molecules and there are molecules which 
contain hundreds of atoms such as proteins, the DNA and RNA extended up to 
several twelves of angstr\"oms. Today are synthetized molecules containing 
until $10^4$ atoms. The human being is the most evolued structure, being 
perhaps the conscience of Universe~\cite{Rosen}. It is composed by 
$10^{28}\simeq 10^5\,\,{\rm mol}$ of atoms, being hydrogen, carbon, nitrogen, 
and oxygen the majority of them. Even if more than one hundred of chemical 
elements have been catalogated in the Mendeleev periodic classification 
just four of them are the essential ones to build up the basic molecular 
structures that allows life and intelligent life.  
\section{The numerical clusters} 
\par
With the fundamental constants $c$, $G$, $\hbar$, $k_{\rm B}$, and the 
electron mass 
\begin{equation}
m_{{\rm e}^-} = 0.510\,999\,06\,\,{\rm MeV}\,c^{-2},
\label{otto}
\end{equation}
where 
$1\,\,{\rm eV} \equiv 1.602\,177\,33\times 10^{-19}$ joule, 
which is the lightest massive elementary fermion among all the elementary 
particles of the electroweak standard model, the mass of the proton 
\begin{equation}
M_{{\rm p}^+} \simeq 938.271\,99\,\,{\rm MeV}\,c^{-2}
\label{novve}
\end{equation}
and of the neutron,
\begin{equation}
M_{{\rm n}^0}\simeq 939.565\,33\,\,{\rm MeV}\,c^{-2}
\label{dieci}
\end{equation}
and the electron electric charge
\begin{equation}
e = - 1.602\,177\,335\times 10^{-19}\,\,{\rm coulomb},
\label{undici}
\end{equation}
we can obtain adimensional numbers such as
\begin{equation}
\frac{M_{{\rm n}^0}}{m_{{\rm e}^-}}\simeq 1839,
\quad 
\frac{M_{{\rm p}^+}}{m_{{\rm e}^-}}\simeq 1836
\label{dodici}
\end{equation}
or even the Sommerfeld fine structure constant
\begin{equation}
\alpha_{\rm em} \equiv \frac{e^2}{\hbar c}\simeq \frac{1}{137} 
\label{tredici}
\end{equation}
as well as the dimensional quantities such as the radius of the Bohr level 
of the Hydrogen atom
\begin{equation}
a_0 \equiv \frac{\hbar}{m_{{\rm e}^-}\,c}\simeq 0.5\times 10^{-8}\,\,{\rm cm}
\equiv 0.5\,{\rm\AA}.
\label{quatord}
\end{equation}
Realize the absence of the universal constant of gravitation $G$ in these 
three last relations which cannot be combined with $\hbar$ and $c$ to form 
a gravitational fine structure constant analogous to $\alpha_{\rm em}$. 
The characteristic ondulatory length of the electron 
at velocities near to that of the light $c$, by its turn, is given by the 
reduced Compton wavelength 
\begin{equation}  
\lambdabar_{{\rm e}^-}\equiv\frac{\lambda_{{\rm e}^-}}{2\pi} =
\frac{\hbar}{m_{{\rm e}^-}\,c} = 
\alpha_{\rm em}\,a_0
\simeq\frac{1}{137}\,a_0
\label{quindci}
\end{equation}
and when all energy of an electron, $m_{{\rm e}^-} c^2$, 
is electrostatic potential energy kind, $e^2/a$, then
\begin{equation}
m_{{\rm e}^-}c^2 = \frac{e^2}{a},
\label{sedici}
\end{equation}
resulting the electron classical radius,
\begin{equation} 
a = \frac{e^2}{m_{{\rm e}^-}\,c^2}
\equiv 3\times 10^{-13}\,\,{\rm cm}\simeq 3\,\,{\rm fermi}
\label{diciasette}
\end{equation}
\par
When we build up numbers involving the constant $G$ it is possible to obtain 
numbers that are centered in a cluster of the order of $10^{40}$ as 
the ratio between the forces of the electrostatic and gravitational 
interactions among a proton and an electron,
\begin{equation}
\frac{F_{\rm electr}}{F_{\rm grav}} = \frac{e^2}{{G\, M_{{\rm p}^+}}\,
m_{{\rm e}^-}}
\simeq 10^{40}
\label{diciotto}
\end{equation}
which is much larger than the Avogadro constant, 
$N_A = 6.022\,141\,99(47)\times 10^{23}$ mol$^{-1}$. In the atomic scale, 
electric forces are dominant but in large scale such forces anulate each other
and the gravitation becomes dominant. If a nucleon with the mass $M_n$ of 
almost 1 GeV is a black hole in 
rotation, of Schwarzschild~\cite{SWGrav}, its radius is obtained 
by equalling the 
kinetic energy  $\frac{1}{2} m v^2$, to the gravitational potential energy 
$G\,m\,M_n/r$. Considering the limit $v = c$, the Schwarzschild radius is
\begin{equation}   
r_S = 2 G\frac{M_n}{c^2},
\label{dicianove}
\end{equation}
corresponding to the gravitational lenght of a nucleon 
\begin{equation}
a_g = G\frac{M_n}{c^2}\simeq 10^{-52}\,\,{\rm cm}
\label{venti}
\end{equation}
in which we do not take into account the factor 2. If a nucleon of classical
radius $a\equiv e^2/(m_{{\rm e}^-}\,c^2)$ reaches the gravitational lenght 
$a_g$, the ratio between these lengths will be
\begin{equation}
\frac{a}{a_g} = \frac{e^2}{G\,M_n\,m_{{\rm e}^-}}\simeq 0.2\times 10^{40}
\label{ventuno}
\end{equation}
that shows that a nucleon is much larger when compared to what it would 
be if it were collapsed in a Schwarzschild black hole. We 
classify two adimensional numerical groups. The first one near to the unit,
such as ${m_{{\rm e}^-}}/{M_n}\simeq 1/1836$ 
or $e^2/\hbar\,c\simeq 1/137$ and their inverses. Another cluster of 
adimensional numbers involve the 
constant of gravitation $G$, centered around $10^{40}$ as 
\begin{mathletters}
\begin{eqnarray}
\frac{e^2}{G M^2_{n}} & = & \frac{1}{1836}\times (0.2\times 10^{40}), \\
\label{ventquauno}
\frac{\hbar\,c}{G M^2_n} & = & \frac{137}{1836}\times (0.2\times 10^{40}), \\
\label{ventquadue}
\frac{e^2}{G M_n m_{{\rm e}^-}} & = & 1\times (0.2\times 10^{40}), \\
\label{vntqutre}
\frac{\hbar\,c}{G\,M_n\,m_{{\rm e}^-}} & = & 137\times (0.2\times 10^{40}), \\
\label{vntqquat}
\frac{e^2}{G\,m^2_{{\rm e}^-}} & = & 1836\times (0.2\times 10^{40})
\label{vntiququat}
\end{eqnarray}
\end{mathletters}
always proportional to $G^{-1}$. Let us denotate this cluster of numbers by
${\cal N}_1$.
\section{The change of universal physical constants}
\par
With the scrutiny of the numerical clusters we have, this way, a cluster 
centered in ${\cal N}_0\sim 1$ being this last one centered in 
${\cal N}_1\sim 10^{40}$. 
Considering the time variation\cite{ETeller} 
of the universal physical constants, 
the time functional behavior of the gravitation universal constant 
$G=G(t)\sim t^{-1}$, suggested for the first time 
by Dirac,~\cite{Dirac,Dicke}  
it is possible to verify, sistematically, in this case that
the unit group ${\cal N}_0\sim 1$ is not affected. However, we verify that
all the elements of the cluster ${\cal N}_1$ are shifted beneath 
${\cal N}_2\sim 10^{40}$. It is also considered the variation of the 
elementary electric charge\footnote{As a matter of fact, the quantum 
of electric charge is the charge of the down-like quark flavors 
with $-\frac{1}{3}|e|$ for a particle state.} $e$, in such a manner that it 
diminishes with the 
expansion of the Universe. It is interesting to observe that the elements 
of the unit group ${\cal N}_0$ as well as the ones of the cluster 
${\cal N}_1$ will spread-out.
\section{The High Energy Physics and the Cosmology} 
\par
On dimensional grounds it can be established the Planck scale, inaccessible 
to any conceivable experiment,\\
Energy:
\begin{mathletters}
\begin{equation}
E_{\rm Planck} = \left (\hbar c^5/G\right )^\frac{1}{2}\simeq 
1.2\times 10^{19}\,\,{\rm GeV},
\label{eum}
\end{equation}
Mass:
\begin{equation}
M_{\rm Planck} = \left (\hbar c/G\right )^\frac{1}{2}\simeq
2.1\times 10^{-8}\,\,{\rm kg},
\label{edois}
\end{equation}
Time:
\begin{equation}
t_{\rm Planck} = \left (\hbar G/c^5\right )^\frac{1}{2}\simeq
5.4\times 10^{-44}\,\,{\rm sec},
\label{etres}
\end{equation}
Length:
\begin{equation}
l_{\rm Planck} = \left (\hbar G/c^3 \right )^\frac{1}{2}\simeq
1.6\times 10^{-35}\,\,{\rm m},
\label{etresex}
\end{equation}
Density:
\begin{equation}
\rho_{\rm Planck} = \frac{c^5}{\hbar G^2} \simeq 
5.1\times 10^{96}\,\,{\rm kg}\,\,{\rm m}^{-3} 
\label{equatr}
\end{equation}
\end{mathletters}
involving the $c$, $G$, and $\hbar$ universal constants. In natural units, 
the Newtonian gravitational constant $G$ is associated with the Planck 
energy scale according to
\begin{equation} 
\frac{1}{\sqrt G} = E_{\rm Planck}\approx 10^{19}\,\,{\rm GeV}.
\label{eecinq}
\end{equation}
Gravitation is closely connected to the Planck scale since $E_{\rm Planck}$ 
is the characteristic scale of gravity quantization~\cite{Polchinski}. 
However, the Planck scale is invisible to our highly selective 
detectors which are looking up to the ${\rm kTeV}$ or ${\rm PeV}$ 
scale,~\cite{fnal} with the prefix `peta,' ${\rm P}\equiv 10^{15}$, 
and very  beyond in the Cosmic Ray Physics~\cite{Auger,Ginzburg} on the 
way of a  systematic phenomenology to find new physics. 
The horizon of the observable Universe is determined by the 
Hubble length, 
\begin{equation}
L_{\rm Hubble} = \frac{c}{H_0} 
\label{vntquex}
\end{equation}
where the universal constant 
$H_0 = 100\,h_0\,\,{\rm km}\,\,{\rm s}^{-1}\,\,{\rm Mpc}^{-1}$ is the Hubble 
present expansion rate of the Universe with the adimensional ignorance 
parameter~\cite{PDG2000} 
$h_0 = (0.71\pm 0.07)\times^{1.15}_{0.95}$ and 
$1\,\,{\rm pc}\,\,({\rm parsec})\simeq 3.262\,{\rm ly}
\simeq 3.086\times 10^{16}\,\,{\rm m}$. 
In a energy natural unit 
\begin{equation}
H_0 = 2.13\,h_0\times 10^{-33}\,\,{\rm eV}.
\label{venqpr}
\end{equation}
From the Planck time scale $t_{\rm Planck}\simeq 10^{-44}\,\,{\rm sec}$ 
up to the Hubble time~\cite{Hubble}
\begin{equation}
t_{\rm Hubble} = H_0^{-1}\approx 10^{10}\,\,{\rm yr}\simeq 10^{17}\,\,{\rm sec}
\label{alaltr}
\end{equation}
there are $61$ orders of magnitude, the same as the spread of time in seconds 
to be scrutinyized since the Big Bang to the age of the Universe. 
Taking the value $L_{\rm Hubble}\simeq 15\times 10^9\,\,{\rm ly}$, 
in terms of the electron classical radius, $a$, there is a 
third cluster,
\begin{equation}
{\cal N}_2\equiv\frac{L_{\rm Hubble}}{a}\simeq 5\times 10^{40}
\label{vencinque}
\end{equation}
coincident with the ${\cal N}_1$ cluster. Such coincidence, 
${\cal N}_1 = {\cal N}_2$, seems to indicate an unknown deep connection 
between the universal 
physical constants and the structure of the Universe. Perhaps such 
large numbers realize the  
${\cal N}_1 = {\cal N}_2$ coincidence not so occasionally. 
The establishment of the correlation
$$
{\rm numerical\,\, cluster}\,\,{\cal N}_1\longleftrightarrow  
{\rm elementary\,\,particles}
$$
and
$$
{\rm numerical\,\,cluster}\,\, 
{\cal N}_2\longleftrightarrow {\rm large\,\,scale\,\,Universe},
$$
accordingly to the Dirac large numbers hypothesis, the ${\cal N}_1$ 
and ${\cal N}_2$ clusters ought to be connected by a simple 
mathematical relation. Particularly, 
\begin{equation}
{\cal N}_1 = {\cal N}_2 \equiv {\cal N}.
\label{vensei}
\end{equation}
Thus, it is possible to verify the existence of a numerical series, 
\begin{equation}
{\cal N}^0,\,{\cal N}^{\pm\frac{1}{2}},\,{\cal N}^{\pm 1},\,
{\cal N}^{\pm\frac{3}{2}},\,
{\cal N}^{\pm 2},
\label{vensttte}
\end{equation}
i.e., ${\cal N}^{\pm n/2}$ being $n=0,1,2,3,4$, only, all along for the 
Universe, with ${\cal N}$ contained in the interval 
$10^{38}\leq{\cal N}\leq 10^{42}$.  
Some numbers of such a series can be easily interpreted such, for instance, 
\begin{equation} 
            \frac{{\rm density\,\,of\,\,a\,\, nucleon}}
            {{\rm density\,\,of\,\,the\,\,Universe}} = \frac{{\cal N}^2_2}
            {{\cal N}_1} = {\cal N}
\label{ventoto}
\end{equation}
to the cluster ${\cal N}$, 
\begin{equation}
\frac{{\rm Compton\,\,length\,\,of\,\,a\,\,nucleon}}{{\rm Planck\,\,length }}
= {\cal N}^{\frac{1}{2}}_1,
\label{ventottdue}
\end{equation}
or
\begin{equation}
\frac{{\rm Planck\,\,mass}}{{\rm nucleon\,\,mass}}
= {\cal N}^{\frac{1}{2}}_1,
\label{ventottre}
\end{equation}
for ${\cal N}^\frac{1}{2}$, and
\begin{equation}
\frac{{\rm mass \,\,of\,\,the\,\,Universe}}{{\rm Planck\,\,mass}} = 
{\cal N}^\frac{1}{2}_1\times {\cal N}_2 = {\cal N}^\frac{3}{2},
\label{ventnove}
\end{equation}
or still, 
\begin{equation}
\frac{{\rm extension\,\,of\,\,the\,\,Universe}}{{\rm Planck\,\,lenght}}
= {\cal N}^\frac{1}{2}_1\times {\cal N}_2 = {\cal N}^\frac{3}{2}
\label{trenta}
\end{equation}
for the cluster ${\cal N}^\frac{3}{2}$, and finally,
\begin{eqnarray}
{\rm total\,\,number\,\,of\,\,nucleons\,\,in\,\,the\,\,Universe} & = & 
\nonumber\\
{\cal N}_1\times {\cal N}_2 = {\cal N}^2 
& \approx & 10^{80}
\label{trnttuno}
\end{eqnarray}
which is the Eddington number, the largest number of the 
whole physics together with the total number of photons, $10^{89}$.
\section{Conclusions}
\par
The simple but meaningful algebraic combinations of the 
universal physical constants are grouped in regular numerical
clusters building up a finite geometrical series with a step of $10^{20}$ 
and oposite extremes $10^{\pm 80}$. 
Introducing the system of natural units 
in which $[c]=[\hbar]=[k_{\rm B}]=1$, the High Energy Physics allows a 
one-dimensional system of units. 
Such inevitable one-dimensionality is the natural direct result of the 
Universe ultimate constituents search, according to Georgi ``our pride 
and our curse''~\cite{HGeorgi}.  
Admitting the inverse time variation 
of the gravitation universal constant, the whole cluster of numbers
centered around the unit is not afected, even though the cluster
centered in $10^{40}$ is deviated below this scale. 
All clusters spread-out with the elementary electric charge reduction. 
\par
All surprises and the new physics islands must be in some specific 
places on a natural 
unit onedimensional scale. Such islands contain an apparently inexhaustible 
source of new physics~\cite{end} such as the continuous constatations of the 
deterministic chaos\cite{Kanter} due to a hypersensible initial conditions 
dependence of the classical dynamical system, as well as the cellular 
automata possibility\cite{SW} and the uncertainty principle\cite{WH} even 
for classical systems\cite{Kanter} due to an irreducible limitation on the 
knowledge of the properties of the physical system.
\acknowledgments
F.P. should like to express his gratitude to Professor J. Polchinski for 
your short but very elucidative reply to a question which has motivated 
the scrutiny contained here. N.O.R. is grateful to the CNPq (Brazil) 
for financial support.

\end{document}